\documentclass[11pt,english]{article}
\usepackage{rotating}
\usepackage{epsfig}
\usepackage{latexsym}
\usepackage{graphicx}
\usepackage{psfrag}
\usepackage{amsmath,amssymb}

\makeatletter
\def\section{\@startsection {section}{1}{\z@}{-3.5ex plus -1ex minus
 -.2ex}{2.3ex plus .2ex}{\large\bf}}
\def\subsection{\@startsection{subsection}{2}{\z@}{-3.25ex plus -1ex
minus -.2ex}{1.5ex plus .2ex}{\normalsize\bf}}
\makeatother
\makeatletter
\def\theequation{\arabic{section}.\arabic{equation}}

\@addtoreset{equation}{section}
\renewcommand{\theequation}{\thesection.\arabic{equation}}
\makeatother

\newcommand{\captionfonts}{\small}
\makeatletter  
\long\def\@makecaption#1#2{%
  \vskip\abovecaptionskip
  \sbox\@tempboxa{{\captionfonts #1: #2}}%
  \ifdim \wd\@tempboxa >\hsize
    {\captionfonts #1: #2\par}
  \else
    \hbox to\hsize{\hfil\box\@tempboxa\hfil}%
  \fi
  \vskip\belowcaptionskip}

\newcommand{\gesim}{\,\raisebox{-.3ex}{$_{\textstyle >}\atop^{\textstyle\sim}$}\,}
\makeatother   

\catcode`@=11
\def\marginnote#1{}
\newcount\hour
\newcount\minute
\newtoks\amorpm
\hour=\time\divide\hour
by60
\minute=\time{\multiply\hour by60 \global\advance\minute
by-\hour}
\edef\standardtime{{\ifnum\hour<12 \global\amorpm={am}
\else\global\amorpm={pm}\advance\hour by-12 \fi
 \ifnum\hour=0
\hour=12 \fi
 \number\hour:\ifnum\minute<10
0\fi\number\minute\the\amorpm}}
\edef\militarytime{\number\hour:\ifnum\minute<10
0\fi\number\minute}
\def\draftlabel#1{{\@bsphack\if@filesw
{\let\thepage\relax
 \xdef\@gtempa{\write\@auxout{\string
\newlabel{#1}{{\@currentlabel}{\thepage}}}}}\@gtempa
 \if@nobreak
\ifvmode\nobreak\fi\fi\fi\@esphack}
\gdef\@eqnlabel{#1}}
\def\@eqnlabel{}
\def\@vacuum{}
\def\draftmarginnote#1{\marginpar{\raggedright\scriptsize\tt#1}}
\def\draft{\oddsidemargin
0.0truein
 \def\@oddfoot{\sl preliminary draft \hfil
\rm\thepage\hfil\sl\today\quad\militarytime}
 \let\@evenfoot\@oddfoot
\overfullrule 3pt
 \let\label=\draftlabel
\let\marginnote=\draftmarginnote
\def\@eqnnum{(\theequation)\rlap{\kern\marginparsep\tt\@eqnlabel}
\global\let\@eqnlabel\@vacuum}
}
\catcode`@=12

\def\bea{\begin{eqnarray}} \def\eea{\end{eqnarray}}
\def\be{\begin{eqnarray}} \def\ee{\end{eqnarray}} 

\newcommand{\tbe}{\tan \beta}
\newcommand{\ctbe}{\cot \beta}
\newcommand{\tm}{\widetilde m}

\newcommand{\str}{{\tilde t_R}}

\newcommand{\C}{\mathcal C}

\begin{document}

\thispagestyle{empty}

\begin{center}
\hfill CERN-PH-TH/2008-197 \\
\hfill UAB-FT-654 \\
\hfill FERMILAB-PUB-08-317-T \\
\hfill ANL-HEP-PR-08-33 \\
\hfill EFI-08-16

\begin{center}

\vspace{1.7cm}

{\LARGE\sc The Baryogenesis Window in the MSSM}

\end{center}

\vspace{1.4cm}

{\sc M. Carena$^{\,a,f,h}$,  G. Nardini$^{\,b}$, 
M. Quir\'os$^{\,b,c,d}$, C.E.M. Wagner$^{\,e,f,g,h}$}\\

\vspace{1.2cm}

${}^a\!\!$
{\em {Fermi National Accelerator Laboratory,
	Batavia, IL 60510, USA }}

${}^b\!\!$
{\em { IFAE, Universitat Aut{\`o}noma de Barcelona,
08193 Bellaterra, Barcelona, Spain}}

${}^c\!\!$
{\em {Instituci\`o Catalana de Recerca i Estudis Avancats (ICREA)}}

${}^d\!\!$ {\em {Theory Division, Physics Department, CERN CH-1211
Geneva 23, Switzerland}}

${}^e\!\!$
{\em {HEP Division, Argonne National Laboratory,Argonne, IL 60439, USA}}

${}^f\!\!$
{\em {Enrico Fermi Institute and 
${}^g\!\!$
Kavli Institute for Cosmological Physics}}\\

${}^h\!\!$
{\em {Physics Department, University of Chicago, Chicago, IL 60637, USA}}

\end{center}

\vspace{0.8cm}

\centerline{\bf Abstract}
\vspace{2 mm}
\begin{quote}\small

Electroweak baryogenesis provides an attractive explanation of the
origin of the matter-antimatter asymmetry that relies on physics at
the weak scale and thus it is testable at present and near future
high-energy physics experiments. Although this scenario may not be
realized within the Standard Model, it can be accommodated within the
MSSM provided there are new CP-violating phases and the lightest stop
mass is smaller than the top-quark mass. In this work we provide an
evaluation of the values of the stop ($m_{\tilde t}$) and Higgs
($m_H$) masses consistent with the requirements of electroweak
baryogenesis based on an analysis that makes use of the
renormalization group improved Higgs and stop potentials, and
including the dominant two-loop effects at high temperature. We find
an allowed window in the $(m_{\tilde t},m_H)$-plane, consistent with
all present experimental data, where there is a strongly first-order
electroweak phase transition and where the electroweak vacuum is
metastable but sufficiently long-lived. In particular we obtain
absolute upper bounds on the Higgs and stop masses, $m_H\lesssim 127$
GeV and $m_{\tilde t}\lesssim 120$ GeV, implying that this scenario
will be probed at the LHC.

\end{quote}

\vfill

\newpage
\section{\sc Introduction}
\label{introduction}

The minimal supersymmetric extension of the Standard Model (MSSM) has
become the preferred candidate for the Standard Model (SM) ultraviolet
completion beyond the TeV scale. The MSSM provides a well defined and
consistent perturbative framework which may be extended up to a high
(GUT or Planck) energy scale.  Among its main virtues, on top of
solving the SM hierarchy problem, the MSSM is consistent with
precision electroweak data, it leads to a natural unification of the
three gauge couplings, and provides a natural candidate for the Dark
Matter of the Universe, namely the lightest neutralino.  The search
for supersymmetric particles is therefore one of the main experimental
goals at the forthcoming Large Hadron Collider (LHC) at CERN.

On the other hand electroweak
baryogenesis~\cite{baryogenesis}-\cite{first} is a very elegant
mechanism for generating the Baryon Asymmetry of the Universe (BAU)
that relies on physics at the weak scale and can therefore be tested
at present accelerator energies, in particular at the Tevatron and the
LHC. It has been shown that electroweak baryogenesis cannot be
realized within the Standard Model framework~\cite{AndH}-\cite{CPSM},
and it is neither feasible in the MSSM for arbitrary values of its
parameters~\cite{early}-\cite{mariano2}. A particular region in the
space of supersymmetric mass parameters was found in the MSSM, where
electroweak baryogenesis has the potential of being successful, dubbed
under the name of light stop scenario
(LSS)~\cite{CQW}-\cite{Cirigliano:2006dg}.
 
The LSS of the MSSM is characterized by a light stop, with a
predominantly right-handed component and a mass close to, or smaller
than, the top quark mass. All other squarks and sleptons are assumed
to be heavier than a few TeV (for example, and for simplicity,
acquiring a common mass $\tm$) to fulfill the present bounds on the
Higgs mass~\cite{Yao:2006px}. 
Large values of $\tm$ protect the model
against large flavor changing neutral current effects, or unacceptably
large CP violation effects and electric dipole moments, but have the
drawback of reintroducing a hierarchy problem~\footnote{Somewhat
reminiscent of the Split Supersymmetry scenario  proposed in
Refs.~\cite{ArkaniHamed:2004yi} although in our case $\tm$ may be 
only moderately large.}.  On the other hand, the Higgsinos and gauginos
are required to be light in order to trigger the required CP-violating
currents needed for baryogenesis as well as providing valuable
candidate for Dark Matter. Light gauginos and Higgsinos can be
technically natural as a consequence of some partly conserved
$R$-symmetry~\cite{R-symmetry}.

Large values of $\tm$ lead to the subsequent appearance of large
logarithms in the one-loop approximation of the Higgs mass used in our
previous EWBG calculation~\cite{Carena:1995bx}. This demands a new
treatment of the Effective Theory (ET) of the LSS below $\tm$, involving
resummation of large logarithms using renormalization group equation
techniques, which allows the computation of the Higgs mass in a
reliable way~\cite{Carena:2008rt}.  
Furthermore, in
reference~\cite{Carena:2008rt} we study the condition of gauge
coupling unification in the LSS and find that it predicts values of
$\tm$ consistent with those required to fulfill the present LEP bounds
on the lightest CP-even Higgs mass.

In this paper we re-analyze the EWBG capabilities of the LSS in the
context of the effective theory presented in
Ref.~\cite{Carena:2008rt}.
The article is organized as follows. In
section~\ref{lightstop} we review the properties of the LSS in some
detail and briefly summarize the main results of
Ref.~\cite{Carena:2008rt} relevant for the present study.  In
particular we describe the features of the low energy theory in which
the heavy supersymmetric partners of the quarks and leptons (except
for the right-handed stop), as well as the heavy Higgs doublet have
been integrated out. In section~\ref{cosmological} we discuss the
possible cosmological scenarios associated with the phase transition
to electroweak and color symmetry breaking
vacua. Section~\ref{numerical} contains the numerical results of the
parameter space consistent with a sufficiently strong electroweak
phase transition. There we present our results as windows in the
$(m_H,m_{\tilde t})$-plane. We show that the requirement of a strong
first-order phase transition provides absolute upper bounds on the
Higgs and stop masses, $m_H\lesssim 127$ GeV and $m_{\tilde t}\lesssim
120$ GeV, and find that all solutions in these windows
correspond to cases where the electroweak vacuum is metastable.  The
technical details of the effective potentials, in the presence of
Higgs and stop background fields, which serve as the basis for the
numerical results of section~\ref{numerical}, are presented in
appendices~\ref{appHiggs} and \ref{appStop}.  In
section~\ref{analysis} we study the decay rate of the previously
computed metastable electroweak vacua and we show that in all cases
their life-time is larger than the life-time of the Universe at all
temperatures.  We reserve section~\ref{conclusions} for our
conclusions and outlook.

\section{\sc The light stop scenario}
\label{lightstop}

The mechanism of electroweak baryogenesis relies on the possible
generation of BAU at the electroweak phase transition. To ensure the
preservation of the generated baryon asymmetry, the baryon number
violating processes must be out of equilibrium at the nucleation
temperature $T_n$. To achieve this, the rate of baryon number
violating processes, which depends on the ratio of the sphaleron
energy to the critical temperature, must be smaller than the expansion
rate of the Universe. Quantitatively, this leads to the condition
$\phi(T_n)/T_n \geq 1$, namely a sufficiently strong first-order phase
transition~\footnote{We use the convention $\phi(T=0)=v=246.22 $GeV.}.
The strength of the phase transition may be analyzed by means of the
finite temperature effective potential. It can be shown that strictly
speaking the value of $\phi(T_n)/T_n$ is, to a first approximation,
directly proportional to the sum of the cube of the couplings of the
light bosonic particles of the model to the Higgs boson, and inversely
proportional to the quartic Higgs coupling, which is in turn
proportional to the square of the Higgs mass. In the SM the only
bosonic particles which couple in a relevant way to the Higgs field
are the weak gauge bosons, with couplings which are governed by the
corresponding weak gauge couplings. The phase transition strength can
therefore be evaluated leading to an upper bound on the mass of the
Higgs boson about 40~GeV, far below the present LEP lower
bounds~\footnote{The former is a perturbative result.
Non-perturbatively, and for allowed Higgs masses, the phase transition
has been proved to be a continuous crossover~\cite{nonpert}.}.

In the MSSM there are additional bosons with relevant couplings to the
Higgs, namely the superpartners of the top quark. Every stop has six
degrees of freedom and therefore the stops could contribute relevantly
to the phase transition strength leading, for sufficiently light
stops, to a strongly first-order phase transition for masses of the
Higgs allowed by the present LEP bound,
$m_h>114.7$~GeV~\cite{Yao:2006px}.  In practice only the (mainly)
right-handed stop may be light. The heaviest (mainly) left-handed
stop has to acquire a mass above a few TeV to achieve agreement with
electroweak precision tests and to ensure a sufficiently heavy Higgs
boson~\cite{Carena:2008rt} compatible with the LEP 
bounds.  On the other hand this favourable improvement on the phase
transition would be substantially reduced by gluinos in the plasma due
to their potentially large contribution to the effective stop mass at
finite temperature, so that gluinos are usually considered heavy
enough to be decoupled from the thermal bath. In practice, this
implies that the gluino mass should be larger than about 500~GeV.

Another problem for the generation of the BAU within the SM is that
the CP-violating sources are highly suppressed.  Therefore new sources
of CP-violation must be present. In the LSS the CP-violating currents
associated with scalar fields are strongly suppressed and therefore
the relevant sources may only be generated by the chargino and
neutralino currents. The charginos and neutralinos should therefore
remain light in this scenario and there should exist non-negligible
phases between the Higgsino and gaugino mass parameters $\mu$
and $M_i$, respectively.  These phases
have important phenomenological consequences inducing potentially
large electric dipole moments (EDM) of the electron and the neutron at
the one-loop level. The one-loop contributions to the EDM's may be
efficiently suppressed if the first and second generation scalar
particles are heavy enough, with masses larger than about 10~TeV. Even
in the absence of one-loop contributions, two-loop contributions
involving the charginos and the Higgs field would remain
sizeable~\cite{edm}. 
They become, however, smaller for values of the CP-odd Higgs
mass larger than about 1 TeV. Still, 
even for very large values of the CP-odd Higgs mass, there is a
contribution induced by the SM-like Higgs boson which, for 
phases of order one, is only an order
of magnitude below the present experimental bounds on the electron 
electric dipole moment. In the following, we shall
identify the CP-odd Higgs mass with $\tilde{m}$.

Summarizing, the generic spectrum of the LSS is constituted by light
charginos and neutralinos, a light stop, heavy first and second
generation squarks and sleptons and, finally, gluinos much lighter
than the heavy scalars but heavy enough to decouple from the thermal
bath. In order to lead to agreement with precision data the light stop
must be predominantly right-handed, and the left-handed stop should
therefore be heavy in order to ensure a large enough Higgs mass.
Moreover, even if the LSS has no specific requirement about the Higgs
sector, as mentioned above a large splitting between the two Higgs
bosons alleviate the MSSM phenomenological problems related to 
flavor or CP-violating effects, because it mimics
at low energy (LE) the Standard Model Higgs sector.

The LSS spectrum contains then light, weak scale particles, as well as
heavy massive particles, with masses much higher than the EW scale.
On the other hand, EWBG is a mechanism that works at the EW scale,
where heavy particles are decoupled. For that reason it seems
appropriate to make the EWBG analysis in the context of the effective
theory of the LSS that was studied in Ref.~\cite{Carena:2008rt} and
that we hereby review briefly.

In mass--independent subtraction schemes particle decoupling is
usually performed by means of a step-function approximation along with
a run-and-match procedure between the underlying theory and the
effective one below every decoupling scale. In particular we will work
in the $\overline {MS}$-scheme and assume, for simplicity, a common
scale $\tm$ for all heavy particles.  Following this criterion at
renormalization scales $\tau$ lower than $\tm$, at which supersymmetry
is broken, the effective Lagrangian turns out to
be~\cite{Carena:2008rt}
\bea
{\cal L}_{\textrm eff} &=& m^2 H^\dagger H-\frac{\lambda}{2}\left(
H^\dagger H\right)^2 - h_t \left[{\bar q}_L \epsilon H^* t_R \right] +
Y_t \left[\overline{{\tilde H}}_u\epsilon q_L {\tilde t_R}^*
\right] 
\nonumber \\ 
&&-\frac{M_3}{2}\Theta_{\tilde g}\, {\tilde g}^a {\tilde g}^a -
\frac{M_2}{2} {\tilde W}^A
{\tilde W}^A -\frac{M_1}{2} {\tilde B} {\tilde B} -\mu {\tilde
H}_u^T\epsilon {\tilde H}_d  - M_U^2 \,\left|{\tilde t_R}\right|^2
\nonumber \\  && - \sqrt{2} \Theta_{\tilde g} G~ {\tilde t_R} {\tilde g}^a
\overline{T}^a \overline t_R  + \sqrt{2} J~ {\tilde t_R} {\tilde B}
\overline t_R
-\frac{1}{6} K \left|{\tilde t_R}\right|^2 \left|{\tilde t_R}\right|^2
 - Q ~ \left|{\tilde t_R}\right|^2
\left|{H}\right|^2 ~+{h.c.}
\nonumber \\
&& +\frac{H^\dagger}{\sqrt{2}}\left( g_u \sigma^a {\tilde W}^a + g'_u
{\tilde B} \right) {\tilde H}_u +\frac{H^T\epsilon}{\sqrt{2}}\left( -
g_d \sigma^a {\tilde W}^a + g'_d {\tilde B} \right) {\tilde H}_d
+{\textrm {h.c.}} ~ ,
\label{lagreff}
\eea 
where  $m^2$, $M_U^2$ are the Higgs and stop mass parameters, $M_i$,
with $i=1,2,3$ are the masses of the gluinos associated with the hypercharge, 
weak and strong  interactions and $\mu$ is the Higgsino mass parameter.
The gluino decoupling is taken into account by the symbol
$\Theta_{\tilde g}$ which is equal to 1 (0) for $\tau\geq M_3\, (\tau<
M_3)$. The effective couplings 
$h_t$, $Y_t$, $G$, $J$, $K$, $Q$, $g_u$ and
$g_d$  in Eq.~(\ref{lagreff})
are obtained from the RG evolution of their values at the scale $\tilde m$
after applying the appropriate
one--loop matching conditions~\cite{Carena:2008rt}
\bea
\label{matchQ}
Q(\tilde m )-\Delta Q&=&\left(\lambda_t^2 (\tilde m )\sin^2\beta +
      \frac{1}{3}~g'^2(\tm) \cos 2\beta\right) \left(1-\frac{1}{2} 
\Delta Z_Q\right)~,\\
\label{matchLamb}
\lambda(\tilde m ) -\Delta\lambda&=& \frac{g^2(\tilde m )+g^{\prime
      2}(\tilde m)} {4} \cos^22\beta 
\left(1-\frac{1}{2}\Delta Z_\lambda\right)~, \label{lambda}\\
K (\tilde m )-\Delta K &=& \left(g_3^2
        (\tm)+\frac{4}{3}~g'^2(\tm)\right)\left(1-\frac{1}{2}\Delta Z_K\right)
        ,\label{matchK}\\
G (\tilde m )-\Delta G&=&g_3 (\tilde m ) 
\left(1-\frac{1}{2}\Delta Z_G\right)~,\\
h_t(\tilde m )-\Delta {h_t}&=&\lambda_t (\tilde m )\sin\beta
\left(1-\frac{1}{2}\Delta Z_{h_t}\right)~,
\label{matchht}
\\
 Y_t (\tilde m )-\Delta {Y_t}&=&
\lambda_t (\tilde m )
\left(1-\frac{1}{2}\Delta Z_{Y_t}\right)~, \label{matchY}\\
g_u(\tm)=g(\tm)\sin\beta~, && g_d(\tm)=g(\tm)\cos\beta ~,
\label{weak1} \\
g'_u(\tm)=g'(\tm)\sin\beta~, && g'_d(\tm)=g'(\tm)\cos\beta
~,\label{weak2}\\
J (\tilde m )=\frac{2}{3}g'(\tilde m ) ~.&& \label{match}
\eea
The thresholds $\Delta Q$, $\Delta \lambda$, $\Delta K$, $\Delta G$,
$\Delta h_t$, $\Delta Y_t$ and $\Delta Z_i$ are computed at one-loop
considering only the numerically dominant contributions proportional
to the strong gauge coupling $g_3$ and the supersymmetric top Yukawa
coupling $\lambda_t$. In general they are functions of the masses
$m^2, M_U^2, M_3, \mu, \tm$, the supersymmetric trilinear coupling
$A_t$ and the ratio of the two Higgs vacuum expectation values
$\tbe$~\cite{Carena:2008rt}. For instance the most
relevant threshold, which already appears at tree--level, is
\bea
\label{deltaQ}
\Delta Q=- \lambda_t^2 \sin^2\beta \frac{\left|\tilde
A_t\right|^2}{\tilde m^2} ~, \eea
where $\tilde A_t=A_t-\mu \ctbe$.  The relations
(\ref{matchQ})-(\ref{match}) only hold at the decoupling scale
$\tm$. Below the decoupling scale we need to run the effective
couplings following their Renormalization Group Equations (RGE) in the
ET~\cite{Carena:2008rt} down to the EW scale after having crossed the
gluino mass scale $M_3$ at which the gluino, which is the lightest
particle to decouple, is integrated out~\footnote{The gluino gives
rise to new thresholds affecting the RGE and generating
discontinuities in the runnings of the low energy couplings and
masses. We take its decoupling into account following the expressions
obtained in~\cite{Carena:2008rt}.}. Since the RGE--evolution resums
the (possibly large) leading logarithms our procedure renders reliable
the evaluation of the effective couplings and hence the EWBG analysis
also for very large values of $\tm$.

\section{\sc Cosmological scenarios}
\label{cosmological}

Due to the large corrections to the effective stop mass at finite
temperature, a strong enough phase transition may only be obtained for
negative values of the stop mass parameter,
$M_U^2<0$~\cite{CQW}. Thus, at zero temperature there are two minima
of the effective potential in the $(\phi,U)$ plane where $\phi=\langle
H\rangle$ and $U=\langle\tilde t_R\rangle$,be taken into account
located at $(\phi_0,0)$ and $(0,U_0)$ and where the value of the
potential is given by $\langle V_H\rangle$ and $\langle V_U\rangle$,
respectively. At finite temperature these minima should evolve from
the corresponding VEVs $\phi(T)$ and $U(T)$ and their cosmological
evolution will strongly depend on the corresponding nucleation
temperatures, $T_H^n$ and $T_U^n$, determined by the tunneling rate
from the symmetry preserving vacuum to the electroweak breaking or
color breaking vacuum, respectively~\footnote{In order to simplify the
different scenarios presented in this section we will identify here
the temperature at which the phase transition ends with the nucleation
temperature $T^n$. Our results are not affected by this approximation
since a more careful evolution of the phase transitions will be taken
into account in the next sections.}.

There are four possible cosmological scenarios:
\begin{description}
\item[Instability region:] 

When $T_U^n>T_H^n$ and $\langle V_H\rangle> \langle V_U\rangle$ the
transition from the unbroken phase to the color breaking one happens
first and, since the color breaking minimum is deeper than the
electroweak minimum, the system will stay in the color breaking
minimum forever. This region, that we call ``instability region'', is
of course unrealistic.

\item[Two--step phase transition region:] When $T_U^n>T_H^n$ and
$\langle V_H\rangle< \langle V_U\rangle$ the transition to the color
breaking minimum also happens first but, since the electroweak vacuum
is deeper than the color breaking one, the system becomes metastable
at a given temperature. If, at a later stage, there were a tunneling
transition from the color breaking to the electroweak minimum, the
system would supercool and the electroweak phase transition would be
much stronger than naively expected. This process was called
``two-step phase transition'' in Ref.~\cite{Schmidt}. In
Ref.~\cite{Cline:1999wi} it was proven that the last phase transition
never happens, which renders this region unrealistic as well.

\item[Stability region:] When $T_U^n<T_H^n$ and $\langle V_H\rangle<
\langle V_U\rangle$ the electroweak phase transition happens first and
since the electroweak minimum is the true vacuum of the theory this
process gives rise to the usual electroweak phase transition. This
region is called ``stability region'' and will be explored in this
paper. As we will show, due to the present bounds on the Higgs mass,
the electroweak phase transition is too weak in this region for the
mechanism of electroweak baryogenesis to take place.

\item[Metastability region:] When $T_U^n<T_H^n$ and $\langle
V_H\rangle> \langle V_U\rangle$ the electroweak phase transition
happens first but the color breaking minimum is deeper than the
electroweak minimum, which makes the system to be in a metastable
phase. We will call this region ``metastability region'', which will
be the main object of the analysis in the rest of the paper. This
scenario will be proven to be viable if the decay rate of the
electroweak to the color breaking minimum is slower than the expansion
rate of the Universe at the corresponding temperature. 

\end{description}

The analysis of the different cosmological scenarios should be done
with the help of the effective potential at finite temperature
$V(\phi,U;T)$ within the effective theory described in
section~\ref{lightstop}. We have therefore followed the
computation of  the zero temperature 
effective potential improved by the one-loop renormalization group 
equations presented 
in Ref.~\cite{Carena:2008rt} and considered
the thermal contribution to two-loop
order for the $\phi$ and $U$ fields given in appendix~\ref{appHiggs} and 
\ref{appStop}, respectively.  We refer the reader to these
appendices for the analytical details and we report the numerical
results in the next section.

\section{\sc Numerical results}
\label{numerical}

\begin{figure}[h]
\psfrag{X1}[][bl]{$\tbe$} 
\psfrag{Y1}[][l]{$\eta/\eta_{BBN}$}
\vspace{.8cm}
\begin{center}
\epsfig{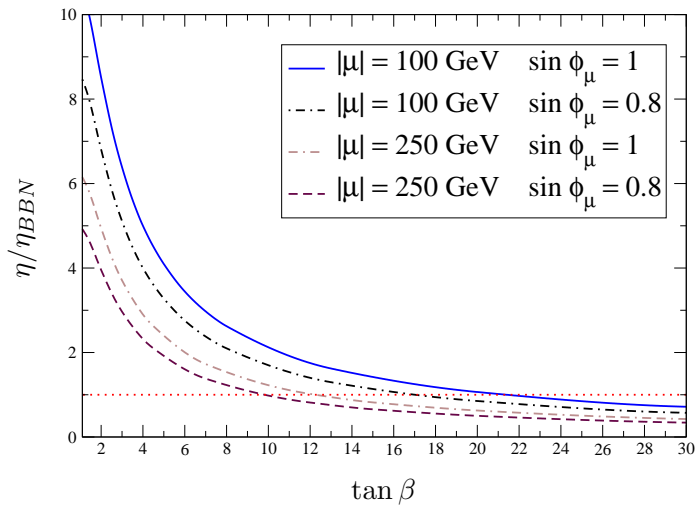}
\end{center}
\caption{$\eta/\eta_{BBN}$ as function of $\tbe$ for several values of
  $\mu$ and imposing $\phi(T_H^n)/T_H^n\simeq 1$, $M_1=M_2=200$ GeV,
  $L_w\simeq 1.7$ and $v_w\simeq 0.1$.} 
\label{eta}
\end{figure}

In this section we will perform the numerical analysis of the phase
transition by using the full effective potential $V(\phi,U;T)$ in the
effective theory of section~\ref{lightstop} evaluated at a value of
the renormalization scale equal to the top-quark pole mass.  We will
search for fundamental parameter combinations satisfying the following
conditions:

\begin{enumerate}
\item
For baryogenesis requirements the relation $\phi(T_H^n)/T_H^n\ge 1$
must be satisfied. In practice, we shall require the condition
$\phi(T_H^c)/T_H^c\ge 0.9$, where the critical temperature 
$T_H^c$ is the temperature at which
the origin and the electroweak minimum at $\phi(T_H^c)$ are
degenerate.  This is a conservative requirement since non-perturbative
results provide in general a stronger first-order phase transition
than perturbative ones~\cite{CK}, and moreover the actual
tunneling temperature $T_H^n$ is smaller than $T_H^c$ and the minimum
$\phi(T)$ increases fast in the small interval $[T_H^c,T_H^n]$.  We
will show in section~\ref{analysis} with an specific example
that $\phi(T_H^c)/T_H^c\ge 0.9$
induces $\phi(T_H^n)/T_H^n>1$ within a good approximation.
\item
We must impose $T_U^n<T_H^f$, where $T_H^f$ is the temperature at
which the electroweak phase transition ends. We thus guarantee that
all the bubbles generated during the phase transition are in
electroweak symmetry breaking vacua and respect either the stability
or metastability conditions.  As will be discussed in
section~\ref{analysis}, the condition $T_U^n<T_H^n$ is fulfilled
whenever the much simpler condition $T_H^c \gtrsim T_U^c+1.6$ GeV is
satisfied. Moreover, we find that there are no stability region points
consistent with the present bounds on the Higgs mass. Moreover, the
metastable vacua satisfying the condition $T_U^n < T_H^f$ do not decay
in the lifetime of the Universe, and therefore, provide a good
realization of the mechanism of electroweak baryogenesis in the MSSM.
\item
The model must be safe from the EDM constraints and generate enough
BAU. This is an important requirement since it is known that the 
generation of BAU requires CP-violating phases of order one. 
One-loop EDM contributions tend to be suppressed
since we have to consider very large values of $\tm$, larger than
about 10~TeV, to overcome the Higgs mass experimental bound and
satisfy the first condition. Also the heavy Higgs sector mass is
identified with $\tm$,  suppressing the two loop contributions
to EDM's~\cite{edm}. 
However, as previously stressed, there are effects associated to the light
SM-like Higgs boson which give contributions which are only one order
of magnitude below the present experimental bounds.
 At a practical level, for smaller values of
$\tm$ the contributions to the EDM's are enhanced at large values of
$\tbe$, and strongly depend on $\tm$ and on the particular choice of
the low energy spectrum.

\item
The successful
generation of the BAU demands moderate or small values of $\tan\beta$.
In practice, there are uncertainties of order one in the theoretical
computation of the BAU. At
the moment however, large variations on the final results appear from
the different
approaches~\cite{Carena:2000id}--\cite{Cirigliano:2006dg} which have
been considered in the literature. The different approaches contain
advantages and/or disadvantages in the treatment of the CPV currents
and sources, the treatment of the diffusion and damping processes and
the possible importance of flavor oscillations. These issues are under
further study and we expect more conclusive results and comparisons
between the different approaches in future publications. 
The leading contributions to BAU decrease as $1/\tan\beta$ for large
values of $\tan\beta$~\cite{Carena:2000id}. 
Therefore, in order to get an
approximate upper bound on the parameter $\tan\beta$ from BAU,
in Fig.~\ref{eta} we plot the ratio of the computed  baryon to entropy
density ratio $\eta$ to the one obtained from Big Bang 
Nucleosythesis~$\eta_{BBN}$. We
use the formalism of Ref.~\cite{Carena:2000id}, 
fixing $\phi(T_H^n)/T_H^n\simeq 1$, $M_1=M_2=200$ GeV,and supposing
bubble walls with width $L_w\simeq 20/T$, velocity $v_w\simeq 0.1$ and
using the definition $\mu=|\mu|e^{i \phi_\mu}$. The results are
only slightly dependent on the stop mass parameters, mainly
through the value of $\phi(T_H^n)/T_H^n$.  
As can be seen Fig.~\ref{eta}, a successful generation of
the baryon asymmetry may be obtained provided $\tbe \lesssim 15$
or, very conservatively, $\tbe \lesssim 5$ (see also Ref.~\cite{Csaba}).
Furthermore, for values of $\tm$ larger than about 10~TeV, 
the generation of the baryon asymmetry may be
obtained without violating the EDM bounds. 
\end{enumerate}

The first two conditions stated above mainly depend on the 
the Higgs quartic coupling $\lambda$, 
the stop quartic coupling $K$ and the stop-Higgs quartic coupling $Q$,
and the Higgs and stop mass parameters. 
This can be intuitively
understood since the barrier developing at finite
temperature strongly depends on $Q$, so that $Q$ is the key parameter
for the first constraint. The remaining parameters determine the
depth of the Higgs and stop tree--level potentials and thus they are
strictly related to the second condition.  
In order to compute the
values of these parameters, one must fix the
parameters $\tm$, $\tilde{A}_t$, and $\tan\beta$, as
well as require the condition of proper electroweak symmetry breaking.
The other free parameters
$Y_t, \mu, M_3, M_2, M_1$ only enter through radiative corrections and
for this reason we simplify the presentation of our analysis by
summarizing the values of the critical low energy parameters as points
on the ($m_h, m_\str$) plane, where
\bea
\label{Mstop}
m_\str^2=M_U^2+\frac{Q}{2}v^2 \qquad \qquad \textrm {with} \qquad
v=246.22 \textrm {~GeV}\, , \eea
and $m_h$ is identified with the second derivative of the one--loop
Higgs effective potential in the ET evaluated at Higgs vacuum 
expectation value
$v$~\cite{Carena:2008rt}. Notice that $m_\str^2$ in Eq.~(\ref{Mstop})
coincides with the (tree--level) stop squared mass in the low--energy
effective theory.

In order to determine the window in which EWBG works we perform a
scanning on the fundamental parameters at the threshold scale $\tm$.
Once we fix $\tm$ the scanning is performed on $A_t, \tbe$ and 
$M_U^2$ since they are the parameters that mostly affect the key effective
couplings.  For the numerical analysis we also have to fix
$\mu,M_1,M_2,M_3$. As we have previously explained we shall demand the
gluino to be sufficiently heavy to be decoupled from the plasma at the
electroweak phase transition. The other parameters are chosen to be at
the weak scale, the phase transition strength being only weakly
dependent on their specific values.

We shall present the results of the numerical analysis for $M_3=500$
GeV and $\mu=M_2=M_1=100$~GeV. Observe that, since we have not
included the weak coupling radiative corrections, the Higgs and stop
potentials are independent of $M_1$ and $M_2$. We shall consider an
uncertainty of about $\pm$3~GeV on our Higgs mass results, reflecting
the lack of weak radiative corrections in the Higgs mass computation,
\begin{figure}[htb]
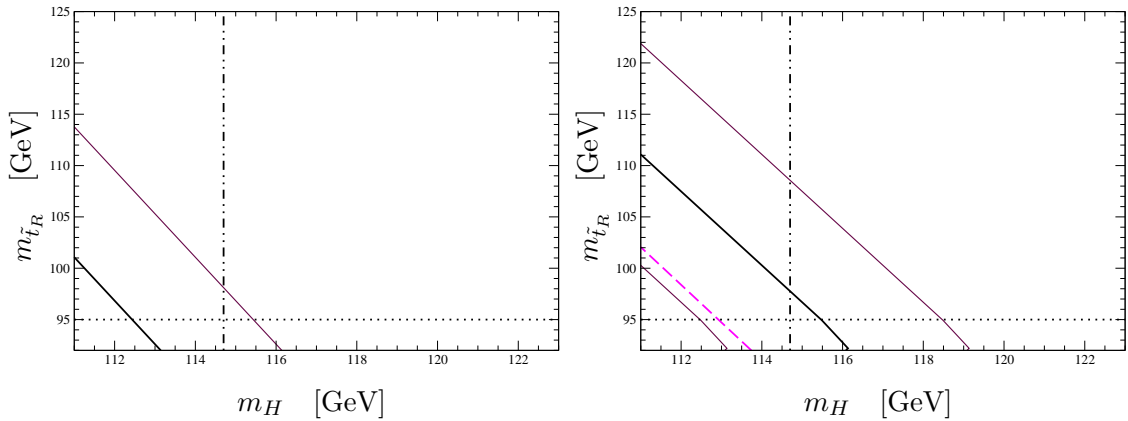

\psfrag{X}[][bl]{$m_H ~~~{\textrm {[GeV]}}$} 
\psfrag{Y}[][l]{$m_\str
~~~{\textrm {[GeV]}}$}
%
\vspace{.8cm}
\begin{center}
\epsfig{file=trans10.eps,width=0.48\textwidth}\hspace{.2cm}
\epsfig{file=trans30.eps,width=0.48\textwidth}
\end{center}
\caption{Window where $\phi(T_H^c)/T_H^c\ge 0.9$ and $T_H^c \ge
T_U^c+1.6$ GeV in the $m_H$-$m_{\tilde t}$ plane for $\tilde m=$ 10
TeV (left panel) and $\tilde m=$ 30 TeV (right panel). 
The allowed
region is below the solid lines and dashed lines for $\tbe \leq 15$
and $\tbe \leq 5$, respectively. 
The thick solid line is obtained
by ignoring the Higgs mass uncertainty, while the solid thin lines
is obtained by including an uncertainty of 3~GeV in the Higgs
mass computation. The Higgs
(stop) mass experimental lower bound is marked by a dotted--dashed (dotted)
line.} 
\label{windows1}
\end{figure}
as well as uncertainties from possible higher-order effects. Under
these conditions the allowed windows for the realization of EWBG in
the MSSM are shown in Fig.~\ref{windows1}, for decoupling scales $\tm=
10$ and $\tm=30$ TeV (left and right panels, respectively).  The right
boundary on each window is provided by the condition $\tbe \le 15$ by
black solid thick line ($\tbe\le 5$ by magenta dashed line, only visible 
in the figure on the right). For the $\tbe
\le 15$ case, we draw three solid lines, corresponding to the bounds
on the Higgs mass obtained by ignoring (black solid thick line), as
well as considering (maroon solid thin lines) the $\pm$~3~GeV
uncertainty on the Higgs mass discussed above.  The allowed area where
the condition $\phi(T_H^c)/T_H^c\ge 0.9$ holds is below (to the left of)
these boundaries. The Higgs and stop mass experimental lower bounds
($m_h>114.7$\,GeV and $m_\str>95$\,GeV~\cite{Yao:2006px}), 
are marked with dotted and dot-dashed lines, respectively.
These results suggest that a heavy
squark spectrum of about 10~TeV may be consistent with electroweak
baryogenesis only for Higgs boson and stop masses at the edge of the
current experimental bounds on these quantities. The situation
improves for 30~TeV, for which an upper bound on the Higgs mass of
about 118~GeV and on the stop mass of about 110~GeV is obtained.

\begin{figure}[htb]
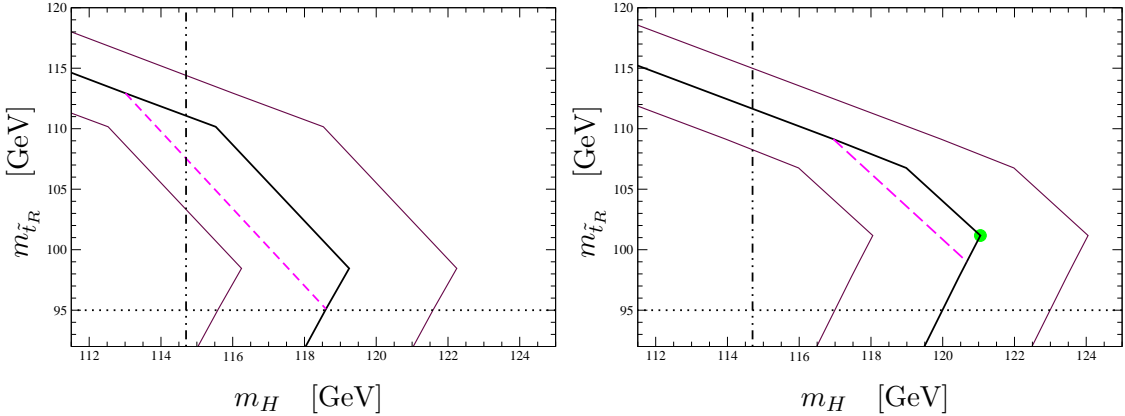

\psfrag{X}[][bl]{$m_H ~~~{\textrm {[GeV]}}$} 
\psfrag{Y}[][l]{$m_\str
~~~{\textrm {[GeV]}}$}
%
\vspace{.8cm}
\begin{center}
\epsfig{file=trans500.eps,width=0.48\textwidth}\hspace{.2cm}
\epsfig{file=trans8000.eps,width=0.48\textwidth}
\end{center}
\caption{Window where $\phi(T_H^c)/T_H^c\ge 0.9$ and $T_H^c \ge
T_U^c+1.6$ GeV in the $m_H$-$m_{\tilde t}$ plane for $\tilde m=$ 500
TeV (left panel) and $\tilde m=$ 8000 TeV (right panel).  The allowed
region is below the solid lines and dashed lines for $\tbe \leq 15$
and $\tbe \leq 5$, respectively. The thick solid line is obtained by
ignoring the Higgs mass uncertainty, while the solid thin lines is
obtained by including an uncertainty of 3~GeV in the Higgs mass
computation.  The Higgs (stop) mass lower bound is marked by a
dotted--dashed (dotted) straight line. In green (right panel) the
point that will be numerically analyzed in the tunneling analysis.}
\label{windows}
\end{figure}

Fig.~\ref{windows} shows similar results for extremal values of the
decoupling scale $\tm = 500$ and $\tm= 8000$~TeV, which are still
compatible with the condition of gauge coupling
unification~\cite{Carena:2008rt}.  The upper almost horizontal border
corresponds to points with $A_t=0$ while going down along the right
border the values of $A_t$ are increasing. The lower boundary
corresponds to the condition $T_H^c \ge T_U^c+1.6$ GeV as trespassing
this boundary we would fall in the instability or two--step phase
transition region.  The allowed area where the condition
$\phi(T_H^c)/T_H^c\ge 0.9$ holds is inside (to the left of) these
solid line boundaries and to the right and above the lines denoting
the stop and Higgs mass experimental bounds, respectively. The stop
and Higgs boson masses can be extended to larger values for these
larger
values of $\tm$, with an upper bound of about 115 and 124~GeV
respectively.

Let us now intuitively understand why the window opens for larger
values of $\tm$.  We first consider, at e.g.~$\tm=500$ TeV, a point
($m_h,m_\str$) just a bit beyond the right central border of the
window. Clearly this point satisfies the first two constraints except
that it exceeds a bit the bound $\tbe\le 15$. Let us now increase
$\tm$ (e.g.~to $\tm=8000$ TeV) without changing the other fundamental
parameters.  As it has been observed in Ref.~\cite{Carena:2008rt}, the
value of $K$ increases for larger values of $\tm$ implying that the
stop tree--level potential becomes less deep and consequently $T_U^c$
decreases. This allows us to consider a larger value of $|M_U^2|$
without loosing the agreement with the second of the requirements
$T_H^f > T_U^n$ ($T_H^c > T_U^c+1.6$ GeV).  Observe that the quartic
coupling of the Higgs (and therefore the Higgs mass) also increases
for larger values of $\tm$, but its change is slower than that of $K$
and therefore increasing $\tm$ affects much less the critical
temperature $T_H^c$ than $T_U^c$.  We have verified that we can then
decrease $A_t$ and $\tbe$ to recover the previous value of
($m_h,m_\str$) [see formulas ~(\ref{deltaQ}) and (\ref{Mstop})] by
keeping the condition $T_H^f > T_U^n$.  Moreover it turns out that by
this procedure the cubic term of the potential is larger, so that the
phase transition is strengthened and the condition
$\phi(T_H^c)/T_H^c\ge 0.9$ can be fulfilled.

\begin{figure}[htb]
\psfrag{mh   mstop}[][]{$m_{\tilde t}$ \quad$m_H \quad \textrm{[GeV]}$}
\psfrag{mtilde}[][]{$\tilde m ~~\textrm{[TeV]}$}
\vspace{.8cm}
\begin{center}
\epsfig{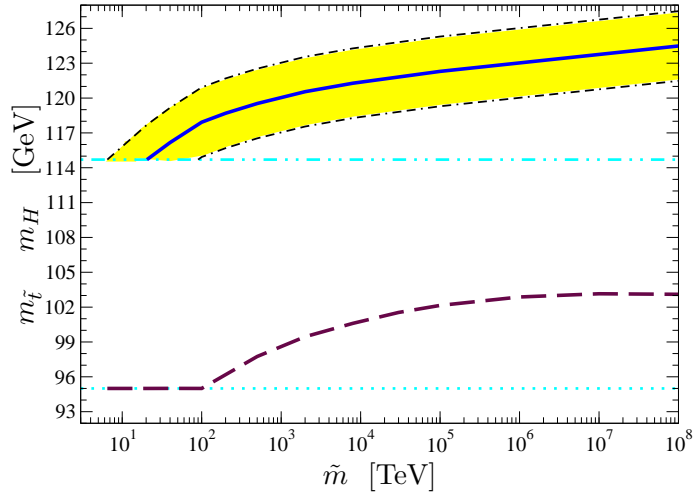}
\end{center}
\caption{$m_H^{max}$ (upper curves) and the corresponding $m_{\tilde
t}$ (lower curve) as functions of $\tilde m$ for $\phi_c/T_c=0.9$ and
$\tan\beta=15$ compatible with their corresponding experimental lower
mass bounds (dotted--dotted--dashed and dotted lines).
\label{maxHiggs}}
\end{figure}

The expansion of the windows is shown in Fig.~\ref{maxHiggs} where the
maximum value of the Higgs mass (solid and dotted-dashed--dashed
lines) [corresponding to the Higgs mass obtained after imposing the
upper bound $\tbe=15$ and the maximum available value of $|M_U^2|$
respecting the condition $T_H^c= T_U^c+1.6$ GeV] and the corresponding
value of the light stop mass (dashed line) are plotted as functions of
$\tm$. The solid lines correspond to the $m_h$ bound obtained by ignoring
the 3~GeV theoretical uncertainty, while
the dotted-dashed-dashed lines correspond the $m_h$ bounds
obtained by considering the theoretical uncertainty.
The dotted--dotted-dashed and dotted lines correspond to the $m_h$ and
$m_{\tilde t}$ experimental mass bounds.  Therefore in
Fig.~\ref{maxHiggs} the minimum value of $\tilde m$ consistent with
EWBG in the MSSM can be extracted and it turns out to be $\tm\gtrsim
6.5$ TeV, while the maximum value of the Higgs mass is about 127 GeV.


A thorough analysis of the effective potential reveals that all points
filling the windows in Figs.~\ref{windows} and \ref{windows1} satisfy
the condition $\langle V_H\rangle>\langle V_U\rangle$.  Therefore they
correspond to metastable electroweak vacua.  For the above region to
be considered as realistic it is necessary to prove that the decay
from the electroweak minimum to the (true) color breaking minimum does
not happen. This requires to compute the probability of tunneling from
the electroweak vacuum to the (deeper) color breaking one. For a point
to be considered realistic this tunneling rate should be smaller than
the expansion rate of the Universe at all temperatures $T\leq T_H^n$.
Due to the similarity between this case and the (inverse) two-step
phase transition scenario where a negative result was obtained in
Ref.~\cite{Cline:1999wi}, we expect this to be the case.  Our
numerical results confirm this fact.

\section{\sc Analysis of the metastability region}
\label{analysis}

In this section we perform the numerical analysis of the transition to
the electroweak breaking phase and the stability of the physical
vacuum. The summary of this analysis is that as stated above, whenever
$T^c_H \gtrsim T_U^c + 1.6$~GeV the electroweak phase transition
happens and ends before the color breaking phase transition and the
system does not decay to the color breaking minimum in one expansion
time of the Universe at any temperature below the nucleation one. We
will illustrate it by analyzing a border--line point in the window for
$\tm=8000$ TeV which corresponds to the maximum allowed value of the
Higgs mass [thick (green) point of Fig.~\ref{windows}].

\subsection{\sc Tunneling from the symmetric phase}
The tunneling probability per unit time and unit volume from the false
(symmetric) to the real (broken) minimum in a thermal bath is given
by~\cite{Linde:1981zj}
\begin{equation}
\frac{\Gamma}{\nu}\sim A(T)\exp{\left[-B(T)\right]},\quad B(T)\equiv
\frac{S_3(T)}{T}
\label{bounce3}
\end{equation}
where the prefactor is $A(T)\simeq T^4$ and $S_3$ is the
three-dimensional effective action. At very high temperature the
bounce solution has $O(3)$ symmetry and the euclidean action is
simplified to
\begin{equation}
S_3=4\pi\int_0^\infty r^2 dr \left[\frac{1}{2} \left(\frac{d \phi}{dr}\right)^2
+V(\phi,T)\right]
\end{equation}
where $r^2=\vec x^2$ and the euclidean equations of motion yield for
the bounce solution the equation
\begin{equation}
\frac{d^2\phi}{dr^2}+\frac{2}{r}\frac{d\phi}{dr}=V^\prime(\phi,T)
\end{equation}
with the boundary conditions $\lim_{r\to\infty}\phi(r)=0$ and
$\left.d\phi/dr\right|_{r=0}=0$.

The nucleation temperature $T^n$ is defined as the temperature at
which the probability for a bubble to be nucleated inside a horizon
volume is of order one and in our case it turns out to happen when
$S_3(T^n)/T^n\sim 135$. Below $T^n$ the transition continues until the
fraction of the causal horizon in the broken phase is of order one,
which can be translated into $S_3(T^f)/T^f\sim 110$ for our
case~\cite{AndH,thomas}~\footnote{We thank Guy Moore for calling our
attention into this conceptual point. At a practical level our windows
do depend weakly on distinguishing $T^n$ from $T^f$ or on the choice
of the numbers used to define them.}.

\begin{figure}[htb]
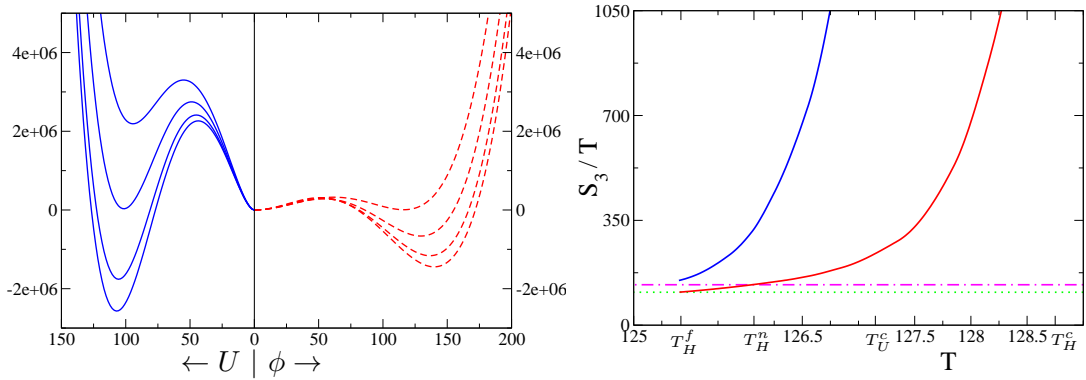

\psfrag{TTT1}[][bl]{}

\psfrag{TTT}[][l]{\hspace{-2mm}$V$  [GeV$^4$] }
\psfrag{X1}[][]{$\hspace{.4mm} \leftarrow U~|~\phi \rightarrow$ }
\psfrag{TnH}[][b]{\tiny{$~T^n_H$}}
\psfrag{TcH}[][b]{\tiny{$~~~T^c_H$}}
\psfrag{TcU}[][b]{\tiny{$~T^c_U$}}
\psfrag{TfH}[][b]{\tiny{$~T^f_H$}}
\vspace{.8cm}
\begin{center}
\epsfig{file=potential.eps,width=0.5\textwidth}
\epsfig{file=action2.eps,width=0.45\textwidth}
\end{center}
\caption{Left panel: effective potential for the Higgs (dashed red
lines in the right plot) and stop (solid blue lines in the left plot)
fields at temperatures (from top to bottom) $T=T_H^c,
T_U^c,T_H^n,T_H^f (128.7,\ 127.1,\ 126.0,\ 125.4$) GeV. Right panel:
bounce actions of tunneling from the symmetric phase towards the
electroweak (dashed red) and colour (solid blue) breaking minima. The
nucleation happens when the action meets the dotted-dashed line and
the transition ends when the action crosses the dotted line.}
\label{potbounces}
\end{figure}

In Fig.~\ref{potbounces} we show the effective potentials (left panel)
along the $\phi$ and $U$ directions at temperatures $T=T_H^c,
T_U^c,T_H^n,T_H^f$, with $T_H^c =128.7$~GeV, $T_U^c =127.1$~GeV,
$T_H^n = 126.0$~GeV and $T_H^f = 125.4$~GeV, for the values of the
supersymmetric parameters yielding the maximum value of the Higgs mass
in the right panel of Fig.~\ref{windows}. The euclidean actions (right
panel) $B_H$ and $B_U$ are computed as function of temperature.  At
$T=T_H^c=128.7$ GeV both actions are infinite. At $T=T_U^c$ the action
$B_U$ is infinite while the action $B_H$ is still too large. At
$T=T_H^n$ the action $B_H\simeq 135$ while $B_U>135$ which means that
the tunneling to the electroweak minimum happens.  At $T=T_H^f$ the
action $B_H\simeq 110$ while $B_U>135$ and therefore our universe
concludes its electroweak phase transition before the beginning of the
colour one.

Notice that in this limiting case the rule $T_H^c=T_U^c+1.6$ GeV is
satisfied and, as anticipated, $T_H^f>T_U^n$. In this particular
example the difference $T_H^f-T_U^n$ is very small because we are
considering a  case in the boundary of the instability region,
but in all other points a larger difference $T_H^f-T_U^n$ is found.

The explicitly considered example also shows that the estimate
$\phi(T_H^c)/T_H^c=0.9$ is a conservative one.  In particular here we
have $T_H^c\approx T_H^n\approx T_H^f$ and
$\langle\phi(T_H^n)\rangle\approx\langle\phi(T_H^f)\rangle$ larger
than $\langle\phi(T_H^c)\rangle$ by $\mathcal O(15\%)$ so that when
$\phi(T_H^c)/T_H^c=0.9$, $\phi(T_H^f)/T_H^f\gtrsim\phi(T_H^n)/T_H^n>1$
({\it i.e.}~all the bubbles generated during the phase transition
produce a strong first-order transition), which seems to be a general
feature in the allowed region.

\subsection{\sc Stability of the electroweak minimum}

Below the temperature $T=T_H^n$ some regions of the universe
are at the electroweak minimum and we must compute the bounce
corresponding to the tunneling to the color breaking minimum,
$B_{HU}$, in order to guarantee the stability of the given point. In
the following we analyze this for the same point of maximal Higgs mass
and $\tm = 8000$~TeV studied in the previous section.

\begin{figure}[htb]
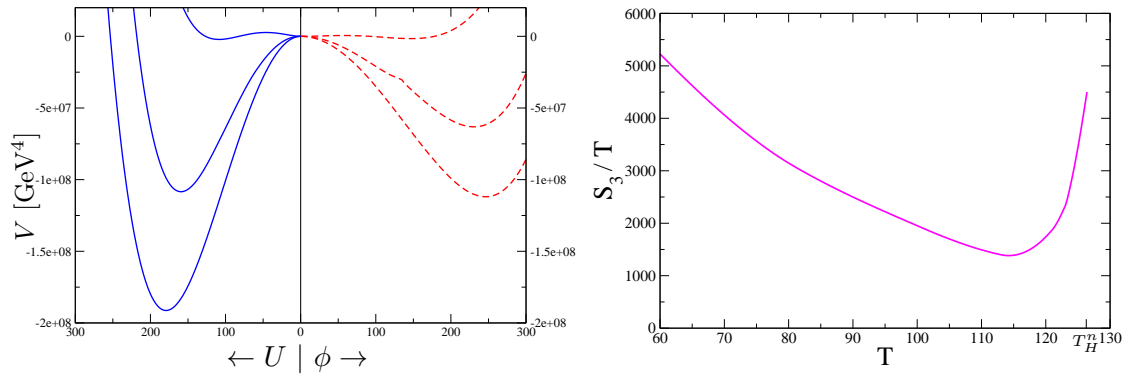

\psfrag{TTT1}[][bl]{}
\psfrag{TTT}[][l]{\small $V$  [GeV$^4$] }
\psfrag{X2}[][]{$\;\qquad \leftarrow U~|~\phi \rightarrow$ }
\psfrag{TnH}[][b]{\tiny{$T^n_H$}}
\vspace{.8cm}
\begin{center}
\epsfig{file=potentialT0.eps,width=0.5\textwidth}
\epsfig{file=actionHU.eps,width=0.47\textwidth}
\end{center}
\caption{Left panel: Effective potential for the Higgs [dashed red
lines in the right plot] and stop [solid blue lines in the left plot]
fields at temperatures $T=126(T_H^n),\ 80,\ 0$ GeV. Right panel:
Bounce action $B_{HU}$ from the electroweak to the colour breaking
minimum as function of $T$.}
\label{potbounce}
\end{figure}

In the left panel of Fig.~\ref{potbounce} we plot the $\phi$ and $U$
potentials for temperatures $T=126,\ 80,\ 0$ GeV and in the right
panel we plot the euclidean action $B_{HU}(T)$. We observe that for
$T=T_H^n=126$ GeV the euclidean action is very large. In fact when
the temperature drops the action $B_{HU}(T)$ drops to a minimum that
nevertheless does not provide a tunneling amplitude that can compete
with the expansion rate of the universe. We have checked that this
effect is even more accentuated in other non-borderline cases.

Finally notice that our results are based on making a path choice for
the bounce $B_{HU}$ that goes through the origin. This path is
consistent with the structure of the minima and the behaviour of the
potential at tree-level~\cite{CQW} and has been considered to be the
proper one to evaluate the transition rate by other similar
analyses~\cite{Schmidt}. For this reason, we believe that our
results are reliable.

\section{\sc Conclusion and outlook}
\label{conclusions}

In this article we have analyzed the strength of the EWPT in the LSS,
which is the most favourable scenario for EWBG in the MSSM. As it was
previously observed in~\cite{Carena:2008rt} the compatibility between
the LSS and the bounds on Higgs and stop masses requires large values
of the soft
supersymmetry breaking masses. Such large values of the soft masses are
also consistent with the condition of gauge coupling unification and
are helpful to suppress dangerous flavor changing neutral current and
CP-violating effects.  Therefore studies on low energy LSS phenomena
are reliable only if they are performed in the effective field theory
where large leading logarithms are resummed. This effective theory was
thoroughly analyzed in Ref.~\cite{Carena:2008rt} and it has been
widely used throughout the present study.

We concentrated on a simple case where all heavy particles (in
particular sfermions -except for the right-handed stop- and the
non-SM-Higgs sector) have a common mass $\tilde m$ while the light
ones (fermions and the right-handed stop) have masses at the
electroweak scale. In the absence of high energy thresholds, gauge
coupling unification predicts values of the scale $\tilde m$ which are
(depending on the precise value of the gluino mass) in the range $\sim
10^{1-3}$ TeV. This range of $\tilde m$ values  has some dependence 
on high energy
threshold effects and/or possible mass splittings at the scale $\tilde
m$.

We have proven that there is a region in the $(\tilde m,m_H,m_{\tilde
t})$ space in which the EWPT is strong enough. The values of $\tilde
m$ are to a large extent in the same  range of values as those 
predicted by gauge
coupling unification. In particular by imposing the LEP bound on the
Higgs mass one obtains a lower bound on $\tilde m> 6.5$ TeV while for
very high values of $\tilde m$ one obtains the absolute upper bound on
the Higgs mass $m_H\lesssim 127$ GeV. As for the stop mass it has to
be light enough in order not to screen the EWPT. Specifically, we
have found in all cases an absolute upper bound on the stop mass as
$m_{\tilde t}\lesssim 120 $ GeV.

As we emphasized in section~\ref{analysis}, in all points of the
allowed BAU windows the electroweak minimum is metastable, while the
true minimum would be one where the color and electromagnetic gauge symmetries
are spontaneously broken. However we have proven that in all cases the
lifetime of tunneling into the color breaking minimum is much larger
than the corresponding age of the Universe and so the electroweak
minimum is stable.

Searches for a light stop and a light Higgs are under way at the
Tevatron collider. The Tevatron can search for a light stop, with mass
below 120~GeV, provided the mass difference between the stop and the
lightest neutralino (assumed to be the lightest supersymmetric
particle) is larger than 30 GeV~\cite{d0excl}--\cite{tevproj}.  For
smaller mass differences, the jets coming from the stop decays are too
soft for the Tevatron experiments to trigger on these events,
rendering the search ineffective.  On the other hand, the existence of
a light Higgs, with mass below 127~GeV, may also be probed at the
Tevatron collider, provided certain sensitivity improvements can be
achieved~\cite{tevproj}.

A light neutralino within the LSS provides a candidate for dark
matter. A proper dark matter relic density may be naturally obtained
in the stop-neutralino coannihilation region, associated with
stop-neutralino mass differences of about 20~GeV~\cite{Csaba}.  The
stop will mostly decay into a charm jet and the light neutralino, but
due to the smallness of the mass difference it will be beyond the
Tevatron reach.  The LHC will be able to provide a definitive test of
the existence of such a light stop: For gluino masses below about
1~TeV, a light stop may be searched for at the LHC in events with
equal sign top-quarks~\cite{ak,Martin:2008aw}.  Even if the gluino
mass is larger than 1~TeV, a light stop may be searched for in events
with high energy jets or photons and missing energy. This latter
search mode, when complemented with Tevatron searches, allows to fully
explore the region of stop masses consistent with electroweak
baryogenesis~\cite{ayres}.  Moreover a light SM-like Higgs may be
searched for at the LHC in different production channels and decay
modes~\cite{pupa}.  Therefore the LHC should be able to provide a
definitive test of this scenario within the next few years.

Before concluding, some comments are worthwhile. First, we have
considered in this paper the case where the MSSM parameter $m_A\simeq
\tilde m$ (and thus the low energy Higgs sector is the SM one) because
it leads to a suppression of 2-loop induced electric dipole moments, 
it requires the smallest values of $\tilde m$ to obtain
a given value of the Higgs mass,
and also because its effective theory is more
tractable. However there is nothing fundamental in considering this
case and one could, following parallel lines to those developed in
Ref.~\cite{Carena:2008rt}, also consider the effective theory with two
Higgs doublets and analyze the corresponding phase transition and BAU,
which favors small values of $m_A$. 
The analysis of such a case, although interesting,
is outside the scope of the present paper.  Second, all phenomena
at low energies, and in particular the EWPT, do depend on the
parameters of the effective theory which, in turn, depend on the
corresponding parameters of the supersymmetric high energy theory. We
have chosen a particular configuration for the latter, 
but other heavy spectra, for instance one in which the
heavy third generation sparticle masses are splitted from the
first and second one, would lead to similar values of the
effective couplings relevant for the EWPT 
and reproduce similar windows.  

Finally, all requirements in the LSS (gauge coupling unification,
consistency with EDM experiments, BAU, actual bounds on the Higgs
mass) lead to values of the supersymmetry breaking parameter ($\tilde
m \gesim 10$ TeV) where the fine-tuning for triggering the electroweak
symmetry breaking is sizeable.  Although this fact can be considered
as a motivation to go beyond the MSSM, still the possibility of
producing the BAU within the MSSM remains as a valid challenge. The
existence of a light SM-like Higgs boson and a stop, with masses below
127~GeV and 120~GeV will be probed at the Tevatron and the forthcoming
LHC experiments and will provide a crucial test of the EWBG scenario
in the MSSM.

%

\subsection*{\sc Acknowledgments}

\noindent 
We would like to thank G.~D.~Moore and T.~Konstandin for useful
comments.  Work supported in part by the European Commission under the
European Union through the Marie Curie Research and Training Networks
``Quest for Unification" (MRTN-CT-2004-503369) and ``UniverseNet"
(MRTN-CT-2006-035863). The work of M.Q. was partly supported by CICYT,
Spain, under contract FPA 2005-02211.  Work at ANL is supported in
part by the US DOE, Div.\ of HEP, Contract DE-AC02-06CH11357. Fermilab
is operated by Fermi Research Alliance, LLC under Contract No.
DE-AC02-07CH11359 with the United States Department of Energy.
  
\section*{\sc Appendix}
\vspace{.25cm}

\appendix
\section{\sc Effective potential for the Higgs field}
\label{appHiggs}

In this appendix we determine the Higgs effective potential at finite
temperature including the leading two-loop corrections in the low
energy effective theory. We focus in the case where heavy enough
gluinos are decoupled from the thermal bath. We will work in the
Landau gauge and in the ${\overline{\rm MS}}$-renormalization scheme.
We will fix the ${\overline{\rm MS}}$-scale $\tau$ to the pole
top-quark mass and consequently all the effective couplings are
evaluated at this scale.

Giving a constant background $\phi_c$ for the real neutral Higgs
boson, the fields of the thermal bath have masses $m_i(\phi_c)$
\bea 
m_W^2=\frac{g^2}{4}\phi_c^2 ~, &&
m_Z^2=\frac{g^2+g'^2}{4}\phi_c^2 ~, 
\nonumber\\
m_h^2=\frac{\lambda}{2}(3\phi_c^2-v^2) ~, &&
m_\chi^2=\frac{\lambda}{2}(\phi_c^2-v^2) ~, 
\nonumber\\
 m_\str^2=M_U^2 + \frac{Q}{2}\phi_c^2 ~, &&
 m_t^2=\frac{h_t^2}{2} \phi_c^2 ~,  \label{stopmass}
\eea
and degrees of freedom 
\begin{eqnarray}
n_{W_L} =  2\,, \qquad n_{W_T}= 4\,, \qquad &&  
n_{Z_L} = 1\,, \qquad n_{Z_T}= 2\,, \nonumber \\
n_{\gamma_L} = 1\,, \qquad n_{\gamma_T}= 2\,, \qquad &&
n_{h} =  1\,, \qquad n_{\chi}= 3\,, \nonumber \\
 n_{t}= -12\,, ~~~~~&&  n_\str=6  \nonumber \ . 
\label{dofSM}
\end{eqnarray}
where the subscript $L (T)$ for gauge bosons is meant for their
longitudinal (transverse) degrees of freedom.  Moreover it is useful
to consider their thermal masses $\overline m_i(\phi_c)$
\begin{eqnarray}
\overline m^2_{Z_L,\gamma_L} &=& 
\frac{1}{2} \left[\frac{1}{4}(g^2+g'^2)\phi_c^2 +\Pi_W +\Pi_B  
\pm  \sqrt{\left( (g^2-g'^2)\frac{\phi_c^2}{4}+\Pi_W-\Pi_B\right)^2
+\frac{1}{4} g^2 g'^2 \phi_c^4}\; \right]
\nonumber\\
&&\qquad \overline m_{W_L}^2  =  m_W^2 + \Pi_W \,, \qquad \qquad
\overline m_h^2  =  m_h^2+ \Pi_h \,,
 \\
&&\qquad \overline m_\str^2   =   m_\str^2 + \Pi_\str \,, \qquad \qquad
\overline m_\chi^2   =  m_\chi^2 + \Pi_{\chi} \,,
\nonumber
\label{masasbar}
\end{eqnarray}
where
\begin{eqnarray}
\Pi_W & = & \frac{7}{3}g^2\; T^2 \,,
\nonumber \\
\Pi_B & = & \frac{22}{9}g'^2\; T^2  \,,
\nonumber \\
\Pi_h & = & \frac{\lambda}{4}\; T^2 
 + \frac{5}{16}g^2\; T^2+\frac{5}{48}g'^2\; T^2+
 \frac{1}{2}h_t^2 \;T^2 \,,
\nonumber\\
\Pi_{\chi} & = & \Pi_h \,,
\nonumber\\
\Pi_{\str} & = & \frac{4}{9} g_s^2\; T^2+ \frac{1}{3} g'^2\;T^2 
           +  \frac{1}{6} Y\;T^2  +  \frac{1}{6} Q\;T^2 \,.
\end{eqnarray}

Considering the Higgs effective potential as a perturbative sum
\begin{equation}
V(\phi_c,T)=V_0+V_1+V_2+\cdots \ ,
\label{potphi}
\end{equation}
where $V_n$ indicates the $n$-th loop potential in the resummed theory
at finite temperature, the tree--level contribution~\footnote{The tree
level Higgs mass is defined such as the one--loop Higgs potential has
a minimum at $v=246.22$ GeV.} is easily obtained by (\ref{lagreff})
and $V_1$ is given by
\bea
 V_1(\phi_c,T)= 
\label{potForm}
 \frac{1}{64 \pi^2} \sum_i
           n_i~m_i^4(\phi_c) \left(
                   \ln \frac{m_i^2(\phi_c)}{\tau^2}
                             -C_i\right)+
                    \sum_i
		        \frac{n_i}{2 \pi^2} J^{(i)} T^4 ~, 
\eea
where $i=W,Z,h,\chi,\str, t$ and $C_W=C_Z=5/6$,
$C_h=C_\chi=C_\str=C_t=3/2$~\footnote{Notice that we are only
considering for simplicity the leading contribution of fields beyond
the SM. The subleading contribution from Higgsinos and/or weak
gauginos would not modify the results in a substantial amount.}.
Since we perform daisy resummation on the $n=0$ modes of the
longitudinal components of the gauge bosons $W_L, Z_L,\gamma_L$ and of
the scalar bosons $h$, $\chi$, $\str$ (no resummation on fermions),
the thermal contributions $J^{(i)}$ are defined by
\begin{equation}
J^{(i)}=
\left\{
\begin{array}{ll}
{\displaystyle J_B(m_i^2)-\frac{\pi}{6}\left(\overline{m}_i^3-m_i^3\right)} &
i=W_L,Z_L,\gamma_L,h,\chi,\str \\ & \\
J_B(m_i^2) & i=W_T,Z_T,\widetilde{T} \\ & \\
J_F(m_i^2) & i=t 
\end{array}
\right.
\label{jotas}
\end{equation}
where the thermal integrals $J_{B,F}$ are
\begin{equation}
J_{B,F}(y^2)=\int_0^\infty dx\; x^2 \log\left(1\mp
e^{-\sqrt{x^2+y^2}}\right)\ . 
\label{termicas}
\end{equation}

We also take into account the logarithmic contributions~\footnote{It
was observed in Ref.~\cite{Carena:1997ki} that non-logarithmic
contributions are negligible in the study of the phase transition.}
coming from the two-loop potential proportional to effective couplings
related to $g_3$ and $\lambda_t$ in their matching condition
(\ref{matchQ})-(\ref{match}). In this approximation the relevant terms
are the sunset diagrams, labeled by $V_{XYZ}$, where $X$, $Y$ and $Z$
are the propagating fields, and figure eight diagrams, labeled by
$V_{XY}$, with propagating $X$ and $Y$ fields.  With this prescription
the two-loop potential turns out to be
\begin{equation}
V_2=V_{\str\str g}+V_{\str\str h}+V_{g\str}+V_{\str h}+
V_{\str\chi}+V_{\str\str}\ ,
\label{dosloops}
\end{equation}
where $g$ stands for gluons and the different contributions are given by

\begin{eqnarray}
V_{\str\str\; g}& = & 
-\frac{g_s^2}{4}(N_c^2-1){\cal D}_{SSV}(\overline m_\str,\overline m_\str,0)
\nonumber \\ 
V_{\str\str\; h}& = & -\frac{1}{2}Q^2 \phi_c^2 T^2 N_c 
              H(\overline m_h,\overline m_\str,\overline m_\str)\; \nonumber\\
V_{g\str}& = &-\frac{g_s^2}{4} (N_c^2-1) {\cal D}_{SV}(\overline m_\str,0) 
\nonumber \\ 
V_{\str\; h}& = & \frac{1}{2}Q^2\sin^2\beta\; N_c\; 
I(\overline m_\str)\,I(\overline m_h)\; 
\nonumber \\
V_{\str\;\chi}& = & \frac{3}{2}Q^2\sin^2\beta\; N_c\; 
I(\overline m_\str)\,I(\overline m_\chi)\; 
\nonumber \\
V_{\str\str}& = & \frac{K}{6}N_c(N_c+1)\; I^2(\overline m_\str))\; 
\label{2looph}
\end{eqnarray}
The functions involved in (\ref{2looph}) are all defined in
Ref.~\cite{twoloop}.

The Higgs potential we have just described is well defined only for
temperatures so large that all squared masses are positive for any
$\phi_c$. When we need to consider lower temperatures, for what
concerns $V_1$ we expand the thermal integrals
$J_{B,F}$~\cite{Dolan:1973qd} and we consider the real part of
$V_1$~\cite{Weinberg:1987vp} while for $V_2$ we use the approximation
\bea 
V_2 \simeq \frac{(\phi_c/T)^2}{32 \pi^2} \left[
\frac{51}{16}g_2^2-3 Q^2+8 g_3^2 Q \log\left(\kappa_H
\frac{T}{|m_\str|}\right) \right] 
\eea
where $\kappa_H\simeq 2.3$.

\section{\sc Effective potential for the stop field}
\label{appStop}

In this appendix we will compute the effective potential at finite
temperature in the background field $U\equiv\str^\alpha u_\alpha$,
where $u_\alpha$ is a constant unit vector in color space which breaks
$SU(3)_c$ into $SU(2)$. We will proceed as in appendix A and present
the result of the two-loop calculation following the same
approximations used for the Higgs potential.

The states contributing to the effective potential are the gauge boson
$B$, four gluons $C$ and the gluon $C'$, five real squarks $\omega$
(would-be Goldstones) and the real squark $\rho$, the Higgs $H$ and
two massive Dirac fermions $f$ coming from the mixing between the
left-handed (third generation) fermion doublet $q_L\equiv q_L^\alpha
u_\alpha$ and the Higgsino, with the corresponding degrees of freedom
\begin{eqnarray}
&&n_{C_L} = 4\,, \qquad n_{C_T} = 8\,, \qquad n_{C'_L}= 1\,, \qquad
n_{C'_T}= 2\,, \nonumber\\
&&n_{B_L} = 1\,,  \qquad n_{B_T} = 2\,,  \qquad n_{H}= 4\,, \qquad
n_{\omega} = 5\,, \\
&& \ \qquad \qquad \qquad n_{\rho}= 1\,, \qquad n_f=-8 \,.  \nonumber
\label{dofStop}
\end{eqnarray}

Their masses in the background $U$ are
\begin{eqnarray}
m_B^2  =  \frac{8}{9}g'^2 U^2 \,,\qquad 
m_C^2  =  \frac{1}{2}g_s^2 U^2 \,,\qquad
m_{C'}^2  =  \frac{2}{3}g_s^2 U^2 \,,\qquad \quad\nonumber\\
m_{\omega}^2  =  M_U^2+\frac{1}{3}K U^2 \,,\qquad
m_{\rho}^2  =  M_U^2+ K U^2 \,,\qquad
m_{H}^2  =  m_h^2 + Q U^2 \,,\nonumber \\
m_f^2  =  \mu^2+Y^2 U^2 \,,\qquad \qquad \qquad \qquad \qquad
\label{Umasses}
\end{eqnarray}
while their thermal masses are defined as $\overline m_i^2 = m_i^2 + \Pi_i$
where $\Pi_\omega=\Pi_\rho=\Pi_\str$ and $\Pi_H=\Pi_h$. 
%
%

The one--loop contribution can be written as expressed in
(\ref{potForm}), where now the index $i$ runs over
$B,C,C',\omega,\rho,H,f$ and the functions $J^{(i)}$ are defined by
\begin{equation}
J^{(i)}=
\left\{
\begin{array}{ll}
{\displaystyle J_B(m_i^2)-\frac{\pi}{6}\left(\overline{m}_i^3-m_i^3\right)} &
i=B_L,\omega,\rho,H \\ & \\
J_B(m_i^2) & i=B_T,C_T,C'_T,{\cal Q} \\ & \\
J_F(m_i^2) & i=f \ .
\end{array}
\right.
\label{jotasU}
\end{equation}
For the functions $J^{C,C'}$ we use their high temperature expansion
except for the contribution of the zero mode (cubic term) which is
screened by the large thermal correction to its mass
$\Pi_{C,C'}=\frac{8}{3}g_s^2 T^2$.

Finally the two-loop diagrams which contribute to $V_2$ are of two
kinds: sunset diagrams labeled by $V_{XYZ}$, where $X$, $Y$ and $Z$
are propagating fields, and figure eight diagrams labeled by $V_{XY}$,
with propagating $X$ and $Y$ fields.  In the following we will denote
$\C\equiv(C,C')$ and $\str\equiv(\omega,\rho)$, $\eta$ being the ghost
fields. Under this prescription $V_2$ is given by

\begin{eqnarray}
V_{\C\C\C} & = & -g_s^2\frac{N_c}{4}\left[ (N_c-2){\cal D}_{VVV} (m_C,m_C,0) +
{\cal D}_{VVV} (m_C,m_C,m_{C'})
\right] \nonumber\\  
& \nonumber\\
V_{\eta\eta\, \C} & = & -g_s^2\frac{N_c}{2} \left[ 2(N_c-1){\cal D}_{\eta\eta
V}(0,0,m_C)+{\cal D}_{\eta\eta V}(0,0,m_{C'})\right] \nonumber \\
& \nonumber\\
V_{\str\str\,\C} & = & -\frac{g_s^2}{4} \left[ (N_c-1){\cal D}_{SSV}
(\overline m_\omega,\overline m_\omega,m_C)+(N_c-1)
{\cal D}_{SSV}(\overline m_\omega,\overline m_\rho,m_C)
\right. \nonumber\\
& + & 
\frac{N_c-1}{N_c}{\cal D}_{SSV}((\overline m_\omega,\overline m_\rho,m_{C'})
+\frac{1}{N_c} {\cal D}_{SSV}
{\cal D}_{SSV}(\overline m_\omega,\overline m_\omega,m_{C'})
\nonumber\\
&+ & \left. N_c(N_c-2)
(\overline m_\omega,\overline m_\omega,0)\right] \nonumber\\ & \nonumber \\
V_{\str\, \C\C} & = &-g_s^2\frac{m_C^2}{8}\left[
(N_c-1){\cal D}_{SSV}(\overline m_\rho,m_C,m_C)
+2\frac{(N_c-1)^2}{N_c^2}{\cal D}_{SSV}(\overline m_\rho,m_{C'},m_{C'})
\right. \nonumber\\
& + & \left. \frac{(N_c-2)^2}{N_c}{\cal D}_{SSV}(\overline m_\omega,m_C,m_{C'})
+
N_c(N_c-2){\cal D}_{SSV}(\overline m_\omega,m_C,0)\right] \nonumber\\ 
& 
\label{2loopU}
\end{eqnarray}
\begin{eqnarray}
V_{GG} & = & -g_s^2\frac{N_c}{8}\left[2(N_c-2){\cal D}_{VV}(0,m_C)
+2{\cal D}_{VV}(m_C,m_{C'})+
(N_c-1){\cal D}_{VV}(m_C,m_C)\right]\nonumber \\ & \nonumber\\
V_{\str\, G} & = & -\frac{g_s^2}{8}\left\{ (N_c-1)
\left[3{\cal D}_{SV}(\overline m_\omega,m_C)+
{\cal D}_{SV}(\overline m_\rho,m_C)\right] \right. \nonumber\\
& + & \frac{1}{N_c}
\left[(N_c+1){\cal D}_{SV}(\overline m_\omega,m_{C'})
+(N_c-1){\cal D}_{SV}(\overline m_\rho,m_{C'})
\right] \nonumber\\
& + & \left. 2N_c(N_c-2)(\overline m_\omega,0) \right\} \nonumber\\ 
& \nonumber\\
V_{\str\str\str} & = & -\frac{K^2}{18}\left[3H(\overline m_\rho,
\overline m_\rho,\overline m_\rho)+
(2N_c-1)H(\overline m_\rho,\overline m_\omega,\overline m_\omega)\right]\, 
T^2\, U^2
\nonumber\\ & \nonumber\\
V_{\str H H} & = &  -2 Q^2 U T^2 H(\overline m_\omega,\overline m_H,
\overline m_H)\; T^2 \nonumber\\ & \nonumber\\
V_{\str\str} & = & \frac{1}{24} K\left[
3I^2(\overline m_\rho)+(4N_c-2)I(\overline m_\rho)I(\overline m_\rho)
+(4N_c^2-1)I^2(\overline m_\omega)
\right] \nonumber\\ & \nonumber\\
V_{\str H} & = & Q\ I(\overline m_H)\left[
I(\overline m_\rho)+(2N_c-1)I(\overline m_\omega)\right] \nonumber
\end{eqnarray}
where all functions involved in (\ref{2loopU}) are defined in
Ref.~\cite{twoloop}.

At low temperatures the potential we have just constructed has problems
as the Higgs effective potential (see appendix A). Also in this case
we extract the real part from the one--loop contribution $V_1$ and we
approximate the two--loop part $V_2$ as follows
\bea
V_2 \simeq \frac{(U/T)^2}{16 \pi^2} 
\left[
  \frac{100}{9}g_2^2-2 Q^2 \log\left(\kappa_U \frac{T}{U}\right)
\right] \,,
\eea
where empirically $\kappa_U\simeq 1.7$.

\end{document}